\documentclass{jfm}
\usepackage{graphicx}
\usepackage{natbib}
\usepackage{epsfig}
\usepackage{epstopdf}

\ifCUPmtlplainloaded \else
  \checkfont{eurm10}
  \iffontfound
    \IfFileExists{upmath.sty}
      {\typeout{^^JFound AMS Euler Roman fonts on the system,
                   using the 'upmath' package.^^J}%
       \usepackage{upmath}}
      {\typeout{^^JFound AMS Euler Roman fonts on the system, but you
                   don't seem to have the}%
       \typeout{'upmath' package installed. jfm.cls can take advantage
                 of  fonts,^^Jif you use 'upmath' package.^^J}%
       \providecommand\mathup[1]{##1}%
       \providecommand\mathbup[1]{##1}%
      }
  \else
    \providecommand\mathup[1]{#1}%
    \providecommand\mathbup[1]{#1}%
  \fi
\fi

\ifCUPmtlplainloaded \else
  \checkfont{msam10}
  \iffontfound
    \IfFileExists{amssymb.sty}
      {\typeout{^^JFound AMS Symbol fonts on the system, using the
                'amssymb' package.^^J}%
       \usepackage{amssymb}%
         \let\leq=\leqslant
         \let\geq=\geqslant
      }{}
  \fi
\fi

\ifCUPmtlplainloaded \else
  \IfFileExists{amsbsy.sty}
    {\typeout{^^JFound the 'amsbsy' package on the system, using it.^^J}%
     \usepackage{amsbsy}}
    {\providecommand\boldsymbol[1]{\mbox{\boldmath $##1$}}}
\fi

\providecommand\bnabla{\boldsymbol{\nabla}}

\newsavebox{\astrutbox}
\sbox{\astrutbox}{\rule[-5pt]{0pt}{20pt}}

\newcommand\p{\ensuremath{\partial}}

\title[Zonal flows driven by librations]{Experimental and numerical study of mean zonal flows generated by librations of a rotating spherical cavity}

\author[A. Sauret, D. C\'ebron, C. Morize, M. Le Bars]%
{A.\ns S\ls A\ls U\ls R\ls E\ls T$^{1}$\footnote{Email adress for correspondance: sauret@irphe.univ-mrs.fr},\ns
D.\ns C\ls \'E\ls B\ls R\ls O\ls N$^{1}$,\ns C.\ns M\ls O\ls R\ls I\ls Z\ls E$^{1,2}$
\and M.\ns L\ls E\ns B\ls A\ls R\ls S$^{1}$\ls}

\affiliation{$^1$ Institut de Recherche sur les Ph\'enom\`enes Hors \'Equilibre, CNRS/Universit\'es Aix-Marseille,\break
49, rue F. Joliot-Curie, BP 146, F-13384 Marseille cedex 13, France \break
$^2$ Laboratoire FAST, UMR 7608, Bat. 502, Campus Universitaire, 91405 Orsay Cedex, France}

\date{20 April 2010}
\volume{000}
\pagerange{\pageref{firstpage}--\pageref{lastpage}}
\doi{S002211200100456X}

\begin{document}

\label{firstpage}
\maketitle

\begin{abstract}
We study both experimentally and numerically the steady zonal flow generated by longitudinal librations of a spherical rotating container. This study follows the recent weakly nonlinear analysis of \cite{busse2010}, developed in the limit of small libration frequency - rotation rate ratio, and large libration frequency - spin-up time product. Using PIV measurements as well as results from axisymmetric numerical simulations, we confirm quantitatively the main features of Busse's analytical solution: the zonal flow takes the form of a retrograde solid body rotation in the fluid interior, which does not depend on the libration frequency nor on the Ekman number, and which varies as the square of the amplitude of excitation. We also report the presence of an unpredicted prograde flow at the equator near the outer wall.
\end{abstract}


\section{Introduction}

Longitudinal librations (reported below as librations) are periodic oscillations of a rotating container about its axis of rotation. Despite the fact that theses oscillations are time-dependent, it has been recently suggested that they can generate non-linearly a steady axisymmetric flow in the liquid interior through the Ekman boundary layer \cite[][]{busse2010}. A better knowledge of this resulting flow is of great interest in geo- and astrophysics \cite[see for instance][]{noir2009} where libration is driven by gravitational interactions and is used to investigate the interior structure of planets \cite[e.g.][]{margot2007,vanhoolst2008}.

Despite of the possible applications, flows driven by libration in rotating containers have not been much studied. \cite{aldridge1969} have observed experimentally that inertial modes can be excited by libration at particular resonance frequencies, which has been confirmed numerically by \cite{rieutord1991}. However in the case of \cite{aldridge1969}, the experimental results are measurements of pressure differences between two points on the axis of rotation, and do not bring information about the resulting flow created in the interior. \cite{tilgner1999} has investigated numerically the linear response to the forcing in the case of a spherical shell and has shown that the presence of an inner core only marginally modifies the resonance frequencies. More recently \cite{noir2009} have studied experimentally by direct flow visualization the presence of centrifugal instabilities in the form of Taylor-G\"ortler vortices near the outer boundary, by varying the frequency and the amplitude of libration. The same group has also performed LDV measurements of libration-driven zonal flows in a librating cylinder \cite[][]{noir2010} in the case of high frequency librations and axisymmetric simulations in a spherical shell \cite[][]{calkins2010}. Finally, a complete weakly non-linear theory of the zonal flow driven by low frequency librations in a sphere has been recently developed by \cite{busse2010} in the absence of direct resonant forcing of any inertial wave. To the best of our knowledge, the main features of this analytical solution have not yet been validated quantitatively. This is the aim of the present work, combining experimental and numerical approaches. The paper is organized as follows. \S 2 gives a brief summary of the governing equations and of the weakly nonlinear analysis of \cite{busse2010}. In \S 3 we present the experimental setup and the numerical model used in this study. Then experimental and numerical results are  compared to the theory in \S 4. Discussion and conclusion are given in \S 5.

\section{Weakly nonlinear theory}

Let us consider a spherical cavity of radius $R$ filled with a homogeneous and incompressible fluid of kinematic viscosity $\nu$. In the inertial frame the cavity rotates with an angular velocity
\begin{equation}
\boldsymbol{\Omega}(t) =\left(\Omega_0+\frac{\Delta \Omega}{2}\cos(\omega_{lib}\,t)\right) \boldsymbol{k},
        \label{eq1}
		\end{equation}		
where $\Omega_0$ is the mean rotation rate, $\Delta \Omega$ is the amplitude of libration, $\omega_{lib}$ the libration frequency and $\boldsymbol{k}$ the unit vector in the direction of the rotation axis. Using respectively $R$ and ${\Omega_0}^{-1}$ as lengthscale and timescale, the dimensionless equations of motion written in the frame rotating at the angular velocity $\Omega_0$ and the sidewall boundary conditions are given by
\begin{subeqnarray}\label{syst:eqn_adim}
  \frac{\p \boldsymbol{u}}{\p t} +\boldsymbol{u}\cdot \bnabla \boldsymbol{u} + 2 \, \boldsymbol{k} \times \boldsymbol{u}& = &
    - \bnabla p + E \bnabla^2 \boldsymbol{u}		\\[3pt]    
  \bnabla \cdot \boldsymbol{u} & = &   0				\\[3pt] 
  \boldsymbol{u} & = & \epsilon\, \boldsymbol{k} \times \boldsymbol{r} \cos(\omega\,t) \quad \mbox{at\ }\quad |\boldsymbol{r}|=1,
\end{subeqnarray}
where $\boldsymbol{r}$ is the spherical radial coordinate, $\boldsymbol{u}$ the velocity measured in the rotating frame, $p$ the modified pressure taking into account centrifugal forces, $E=\nu/\Omega_0 R^2$ the Ekman number, $\epsilon=\Delta \Omega/2\Omega_0$ the normalized amplitude of libration and $\omega=\omega_{lib}/\Omega_0$ the normalized librational frequency. The librational forcing appears in the problem through the boundary condition (\ref{syst:eqn_adim}c). This problem has been recently solved by \cite{busse2010} in the limit:
		\begin{equation}
    		\sqrt{E} \ll \omega \ll \epsilon \ll 1.
        \label{eq_limit}
		\end{equation}
The limit of small Ekman number allows splitting of the velocity field into two parts: a component $\boldsymbol{U}$ describing the inviscid flow in the interior and a boundary layer component ${\boldsymbol{u}}$. Following the weakly non-linear method previously used in the case of precession \cite[][]{busse1968} and tidal forcing \cite[][]{suess1971}, \cite{busse2010} finds an expression for the steady zonal flow in the inviscid interior $\bar{\boldsymbol{U}}$ in the limit of low libration frequencies
\begin{subeqnarray}\label{syst:eqn_sol}
  \boldsymbol{\bar{{U}}} & = & \epsilon^2\,\boldsymbol{k} \times \boldsymbol{r} \, f(|\boldsymbol{k} \times \boldsymbol{r}|^2)		\\[3pt]    
  \mbox{where\ }\quad  f(x^2) & = & \frac{259\,x^2-360}{2400(1-x^2)}.
\end{subeqnarray}
		
The function $f(x^2)$ represents the average difference in angular velocities between the container and the fluid divided by $\epsilon^2$. For $0 \leq r \leq 1$, $f(x^2)$ is negative, i.e. the fluid is expected to rotate in the retrograde direction. Moreover this differential rotation is nearly constant up to $r \sim 0.6$ with a mean value of $-0.154$. The zonal flow can thus be assimilated in the bulk to a retrograde solid body rotation superimposed on the mean rotation, whose amplitude is independent of the libration frequency and Ekman number and changes as $-0.154\,\epsilon^2$. In the present paper, we verify experimentally and numerically these main features. Note that (\ref{syst:eqn_sol}.b) diverges for $x=1$, i.e. near the outer boundary at the equator. Here, the analytical approach requires the introduction of a specific scaling due to the singularity of the Ekman boundary layer \cite[][]{busse2010}. 

\section{Methods}

\subsection{Experimental setup}

\begin{figure}
  \centerline{\includegraphics[height=6cm]{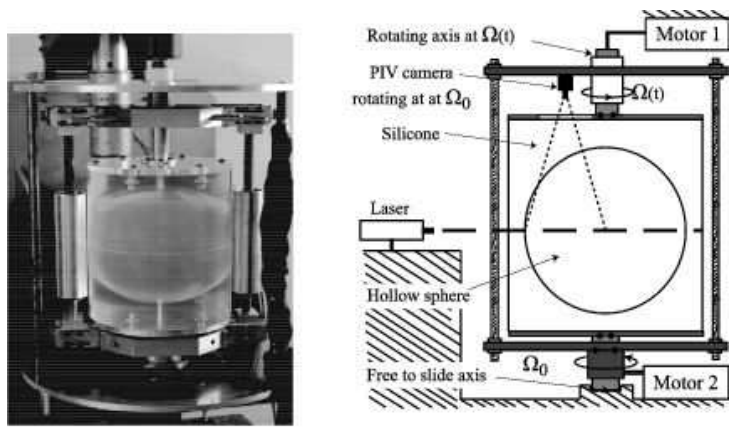}}
  \caption{(a) Photo and (b) sketch of the experimental setup.}\label{fig:schema_setup}
\end{figure}

Figure~\ref{fig:schema_setup} shows a photo and a schematic view of the experimental setup used in this study, which is the same as the one used by \cite{morize2010} to study zonal flows driven by tides. It consists in a hollow sphere, of radius $R=10$ cm, which was molded in a transparent silicone gel to allow flow visualization. The sphere is filled with water and seeded with Optimage particles of $100$ $\mu$m in diameter and of density $1\,$g.cm$^{-3} \pm 2 \%$. The sphere is set in rotation about its vertical axis $(Oz)$ with a mean angular velocity $\Omega_{0}$ up to $85$~r.p.m with a precision of $\pm 0.3\%$. Once a solid body rotation is reached (typically in $\sim 10$ minutes), a librational motion is set using sinusoidal oscillations of the angular velocity of the sphere of the form $\epsilon\,\cos(\omega_{lib}\,t)$ where $\omega_{lib}$ can be chosen between $0.6$ and $120$~r.p.m with a precision of $\pm 0.3\%$. In terms of dimensionless numbers, we have explored the following ranges: the Ekman number $E={\nu}/{\Omega_{0}R^2}\in[10^{-5};10^{-4}$], the ratio between the libration and spin frequency $\omega={\omega_{lib}}/{\Omega_{0}} \in $[$0.04;0.1$] and the amplitude of libration $\epsilon\in $[$0.02;0.15$] with a precision of $\pm 0.6\%$.

In order to measure the velocity field in the equatorial plane induced by librational forcing, we used a rotating particle image velocimetry (PIV) system. A miniature wireless camera $1/4'$ Sharp HighQ CCD $29.4 \times 22$ mm of resolution $576 \times 768$ pixels rotates at a constant angular velocity $\Omega_{0}$ and measurements are made from above through the transparent top surface. The PIV particles are illuminated by a laser sheet, of thickness $3$ mm, produced by a continuous laser ($4$ W) in the equatorial plane. We wait for about $20$ oscillations after turning on the libration forcing to ensure that the response of the fluid is well established, then we start acquiring pictures for PIV measurements using a video transmitter-receptor system. Velocity fields are computed using DPIVSoft \cite[][]{meunier2003} on a $60 \times 80$ grid with a spatial resolution of $3$ mm, close to the laser sheet thickness. We look for the time independent axisymmetric zonal flow induced by the libration whereas the forcing of the sphere is of the form $\epsilon\,\cos(\omega t)$. Velocity fields are thus time-averaged over several periods of libration in order to eliminate the time dependent term. This also significantly enhances the signal to noise ratio. However, our experimental setup only allows collection of PIV data for a limited time. Hence, we cannot set the libration frequency at too low a value because we would not be able to average out the resulting data over enough periods to have a correct velocity profile. In the experiments we have considered frequencies in the range $0.04 \leq \omega \leq 0.1$. Higher and lower libration frequencies have been studied using numerical methods, that we describe in the following section.

\subsection{Numerical approach}

In addition to the experiments, we have performed axisymmetric numerical simulations of the flow within a sphere of radius $R$ in rotation with an angular velocity $\mathbf{\Omega}(t)=\Omega_0(1+\epsilon \cos (\omega t))\mathbf{e_{z}}$. We use a commercial software, Comsol Multyphisics$\copyright$, based on the finite elements method to solve this problem. The numerical grid consists of two domains: (i) a boundary layer domain of thickness $0.035\,R$ all along the outer boundary and the axisymmetric axis, which is discretised in the direction normal to the boundary into $25$ quadrilateral elements with initial thickness of $5.10^{-5}\,R$ and a stretching factor of $1.2$; (ii) a bulk zone with triangular elements. All elements are of standard Lagrange $P1-P2$ type (i.e. linear for the pressure field and quadratic for the velocity field). Note that the finite element method does not induce any particular problem around $r=0$ and that no stabilization technique has been used in this work. The temporal solver is IDA \cite[][]{hindmarsh2005}, based on backward differencing formulas. At each time step the system is solved with the sparse direct linear solver PARDISO\footnote{www.pardiso-project.org}. The number of degrees of freedom (DoF) used in the simulations is constant and equal to $157\,869$ DoF. Our numerical model solves the Navier-Stokes equations in the frame rotating at the velocity $\Omega_0 \, \mathbf{e_{z}}$, with no-slip boundary conditions  and a fluid initially at rest in this frame (i.e. a solid body rotation at $\Omega_0$ in the inertial frame). At time $t=0$, libration of the outer boundary is turned on and computations are pursued until a stationary state is obtained, which is reached typically in less than $10$ libration periods. The velocity is then averaged in time over $5$ libration periods to obtain the steady zonal flow. Results are non-dimensionalised as in the experiment and in the theory. The numerical model has been validated in reproducing the experimental results of \cite{aldridge1969}. In their paper they define a fixed libration Reynolds number $Re_{\omega}~=~{\omega_{lib} R^2}/{\nu}=6.2 \cdot 10^4$ and their applied angular velocity is given by
		\begin{equation}
    		\Omega(t)=\Omega_0+\tilde{\epsilon}\,\omega_{lib}\,\cos(\omega_{lib} t) \quad  \mbox{where $\tilde{\epsilon}={8.0\,\pi}/{180}$ rad}.
		\label{eq_aldridge}
		\end{equation}

\begin{figure}
  \centerline{\includegraphics[width=10cm]{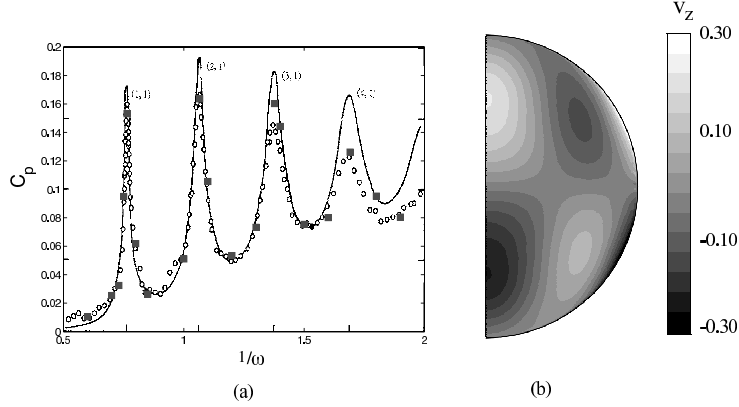}}
  \caption{(a) Time-averaged crest-to-trough amplitude of pressure difference $C_p$ between the center and the pole of the sphere for various frequency ratios $\omega=\omega_{lib} / \Omega_0$ and for a libration forcing  $\Omega_0+\tilde{\epsilon}\,\omega_{lib}\,\cos(\omega_{lib} t)$ with $\tilde{\epsilon}={8.0\,\pi}/{180} rad$ and $Re_{\omega}={\omega_{lib} R^2}/{\nu}=6.2 \cdot 10^4$. The squares stand for our numerical values, the circles for experimental values of \cite{aldridge1969} and the line for the theoretical plot. (b)~Velocity in the $z$-direction (i.e. in the direction of the rotation axis) at time $\omega t= 3 \pi /2\, [2 \pi]$ for the mode $(2,1)$ in the notation of \cite{aldridge1969}, corresponding to $1/\omega=1.066$.}\label{fig:verif_aldridge}
\end{figure}

Pressure measurements from our numerical simulation when systematically changing the libration frequency $\omega$ are presented in figure \ref{fig:verif_aldridge}.a and show an excellent agreement with the experimental results of \cite{aldridge1969} and the numerical results of \cite{rieutord1991}, which validates the numerical model. In figure \ref{fig:verif_aldridge}.a, each peak corresponds to the resonant forcing of a given inertial mode of the rotating sphere and is labeled by two integers $(n,m)$\cite[see][]{aldridge1969}. There is no inertial mode for $| \omega | \geq 2$ and the modes are progressively damped when $1/\omega$ increases due to the reduced coupling between the container's oscillation and the fluid interior as well as to the increased viscous damping as the structure of the forced modes becomes more complex. The velocity in the $z$-direction for the mode $(2,1)$ in the notation of \cite{aldridge1969} is presented in figure \ref{fig:verif_aldridge}.b and shows the inertial wave excited by libration forcing as well as the structure of the outer boundary layer. In the next section, following \cite{busse2010}, we investigate the limit $\omega \ll 1$ where the forcing of inertial modes is negligible, but where a global zonal flow is excited.

\section{Results}

\begin{figure}
  \centerline{\includegraphics[width=13cm]{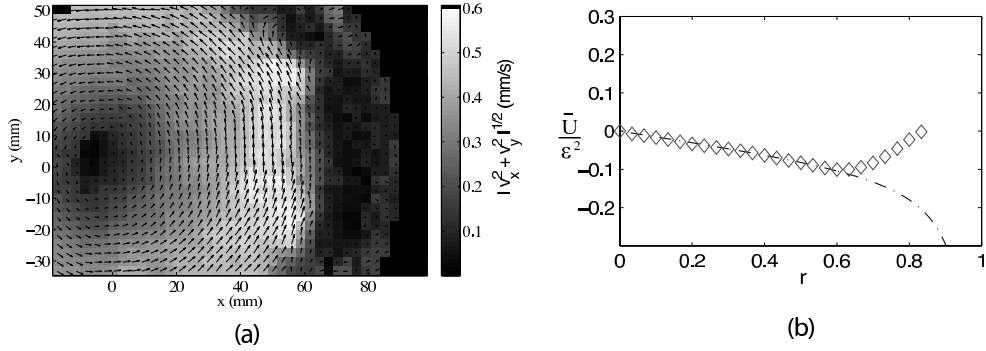}}
  \caption{(a) Time-averaged velocity field obtained by PIV measurement in the equatorial plane for $E=1.15 \times 10^{-5}$, $\epsilon=0.08$ and $\omega=0.1$. The background is colored as the norm of the horizontal velocity. The center of the sphere is at $(0,0)$. (b) Mean experimental dimensionless azimuthal velocity (symbols) corresponding to the velocity field of Fig. \ref{fig:PIV_Field}.a and comparison with the theoretical results of \cite{busse2010} (dashed-dotted line).}\label{fig:PIV_Field}
\end{figure}

\begin{figure}
  \centerline{\includegraphics[width=8cm]{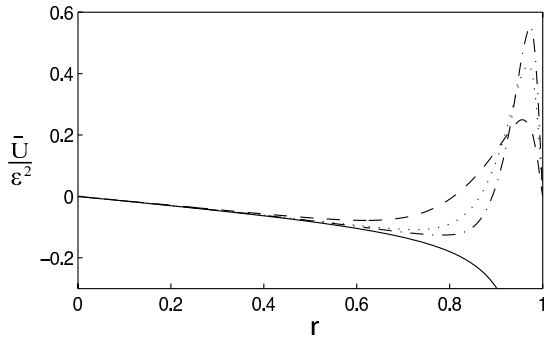}}
  \caption{Dimensionless time-averaged velocity profiles obtained by numerical simulation in the equatorial plane with $E=5 \times 10^{-5}$, $\epsilon=0.2$ and $\omega=0.03$ (dashed line), $\omega=0.06$ (dotted line) and $\omega=0.1$ (dashed-dotted line), compared with the inviscid analytical solution of \cite{busse2010}(continuous line).}\label{fig:numeric_velocity}
\end{figure}

Figure~\ref{fig:PIV_Field}.a shows an example of the velocity field obtained by PIV measurement in the equatorial plane. The stationary flow is azimuthal and axisymmetric. Besides, as the system rotation is clockwise, the zonal flow corresponds here to a retrograde circulation opposed to the rotation of the sphere. From this velocity field we can plot an averaged azimuthal velocity profile at the equator for given parameters in term of dimensionless quantities as a function of the radial distance (Fig.~\ref{fig:PIV_Field}.b). Experimentally, due to optical deformation induced by the planar air-silicone and spheroidal silicone-water interfaces, it is not possible to measure the profile for $r > 0.85$. We observe the steady zonal flow already visible in figure~\ref{fig:PIV_Field}.a. Moreover we can directly compare this dimensionless quantity with the analytical solution given by \cite{busse2010} and we observe an excellent agreement up to $r \sim 0.6$ with no adjustment parameter. An example of the velocity field obtained numerically is shown in figure~\ref{fig:numeric_velocity}, which also exhibits good agreement with the analytical solution in the bulk. For $r\geq 0.6$ a deviation in the prograde direction with respect to the theoritical profile due to the librating outer boundary is observed and will be discussed below. But for now, we concentrate on the mean zonal flow induced in the bulk.

\begin{figure}
  \centerline{\includegraphics[width=8cm]{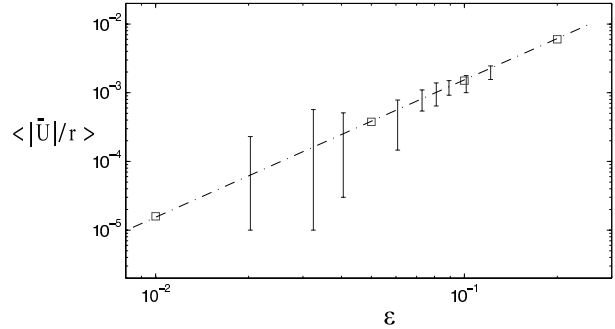}}
  \caption{Dimensionless average amplitude of $|\bar{U}|/r$ between $r=0.1$ and $r=0.6$ as a function of the amplitude $\epsilon$. Experimental results (bars) for $E=2.3 \times 10^{-5}$, $\omega=0.07$, and numerical results (squares) for $E=4 \times 10^{-5}$, $\omega=0.04$ are compared with the theoretical value of \cite{busse2010} (dashed-dotted line).}\label{fig:syst_amplitude}
\end{figure}

We have performed series of experiments and numerical calculations to systematically check the effect of the three control parameters $E$, $\epsilon$ and $\omega$ on this bulk zonal flow. To do so, we define a reproducible method to synthetise the experimental and numerical data. Since the function $f(x^2)$ (\ref{syst:eqn_sol}.b) may be considered as constant up to $r \sim 0.6$, which means that the predicted zonal flow is almost a solid body rotation up to $r \sim 0.6$, we take the average value of the measured $|{\bar{U}|/r}$ between $r=0.1$ and $r=0.6$, i.e. the non-dimensionnalized mean angular velocity, and compare it with the theoretical value $0.154 \epsilon^2$. Note that experimental results are represented by bars (see for example figure \ref{fig:syst_amplitude}) which represent both the uncertainties of the PIV measurements and the deviation of the measured velocity profile from a pure solid body rotation.


In figure \ref{fig:syst_amplitude} we investigate the influence of the amplitude of libration $\epsilon$ on the zonal flow. Experimentally we set $E=2.3 \times 10^{-5}$, $\omega=0.07$ and we systematically change $\epsilon$ between $0.02$ and $0.15$. To explore a larger range of amplitude we have also performed numerical simulations with $E= 4 \times 10^{-5}$, $\omega=0.04$ and $\epsilon\in$[$0.01;0.2$]. 
Both experimental and numerical results are quantitatively compatible with the theory with no adjustment parameter. The steady azimuthal velocity scales as $\epsilon^2$ for a large range of $\epsilon$. When $\epsilon$ becomes larger than $0.2$ the weakly nonlinear hypothesis cannot be used anymore because terms of higher order cannot be neglected. We also notice that even in a range of values where the condition $\omega \ll \epsilon$ in (\ref{eq_limit}) is not fully satisfied, the zonal wind intensity still scales as $\epsilon^2$. In fact, rather than the more restrictive condition (\ref{eq_limit}), we only require that (i) $E \ll 1$ to decouple the bulk and boundary layer flows, (ii) $\sqrt{E} \ll \omega \ll 1$ in order to neglect the excitation of inertial waves \cite[][]{aldridge1969} and to ensure that the spin-up effect of the libration is confined inside the outer boundary layer, and (iii) $\epsilon \ll 1$ to remain in the weakly nonlinear regime.

\begin{figure}
  \centerline{\includegraphics[width=8cm]{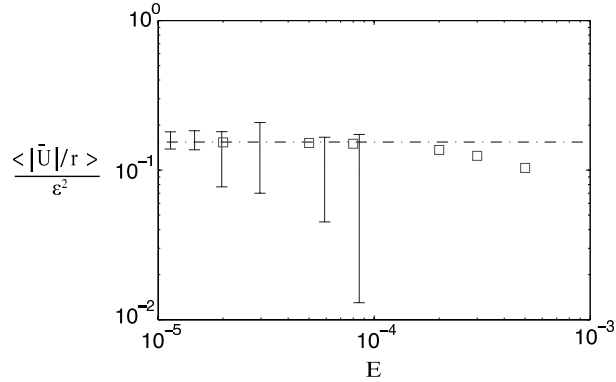}}
  \caption{Dimensionless average amplitude of $|\bar{U}|/r\epsilon^2$ between $r=0.1$ and $r=0.6$ as a function of the Ekman number. Bars are experimental results for $\epsilon=0.08$ and $\omega=0.1$ and squares are numerical results for $\epsilon=0.2$ and $\omega=0.06$. The dashed-dotted line shows the theoretical result of \cite{busse2010}. The velocity profiles have been rescaled by $\epsilon^2$ following the results presented in figure \ref{fig:syst_amplitude}. Experimentally, errorbars increase with the Ekman number mainly due to the fact that our experimental setup does not allow us to average the velocity on a sufficient number of periods when increasing $E$.}\label{fig:syst_ekman}
\end{figure}

In figure \ref{fig:syst_ekman} we report the systematic study of the influence of the Ekman number on the zonal flow. Experimental results are compatible with the no-Ekman dependence predicted by \cite{busse2010} in a large range of Ekman number with no adjustment parameter. This is confirmed numerically up to $E \sim 10^{-3}$. For larger values of the Ekman number, the condition (\ref{eq_limit}) is not fulfilled and we cannot assume that the effect of spin-up is negligible in the bulk. We have also noticed numerically that further decreasing E for a given $\omega=0.1$ (which is not so small) leads to a deviation from the theory. Indeed, forced inertial modes are not expected to be negligible anymore and can perturb the zonal flow. In particular, nonlinear self-interaction of these forced mode can drive localised zonal winds \cite[e.g.][]{morize2010}. These peculiar behavior appears out of the asymptotic limit  (\ref{eq_limit}) under consideration here and will be the subject of a future study.

\begin{figure}
  \centerline{\includegraphics[width=12cm]{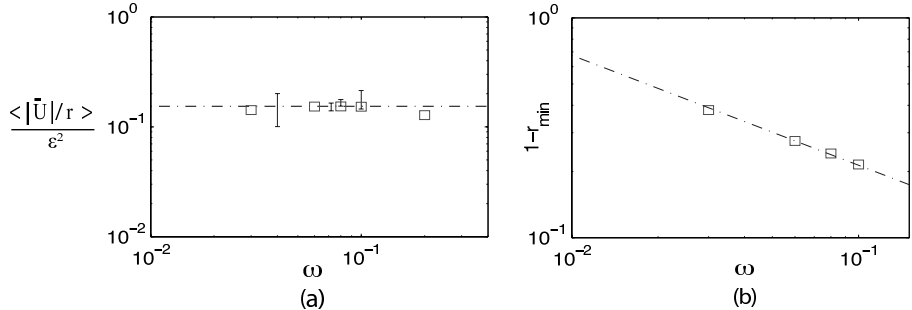}}
  \caption{(a) Dimensionless average amplitude of $|\bar{U}|/r\epsilon^2$ between $r=0.1$ and $r=0.6$ as a function of the frequency of libration $\omega$; squares are numerical values ($E=5 \times 10^{-5}$, $\epsilon=0.2$) and bars are experimental data ($E=1.5 \times 10^{-5}$, $\epsilon=0.1$). The dashed-dotted line shows the theoretical result of \cite{busse2010}. The velocity profiles have been rescaled by $\epsilon^2$ following the results presented in figure \ref{fig:syst_amplitude}. (b) Distance of the minimum of the velocity profile from the outer boundary as a function of $\omega$ obtained by numerical simulation for $E=5 \times 10^{-5}$, $\epsilon=0.2$. The dotted line scales as $1/\sqrt{\omega}$, representative of a skin effect.}\label{fig:graphe_omega}
\end{figure}

Figure \ref{fig:graphe_omega}.a shows that the amplitude of the zonal wind in the bulk does not depend on $\omega$, as suggested by Busse. Nevertheless, as can be seen in figure \ref{fig:numeric_velocity}, the flow near the outer wall changes with $\omega$. We expect this prograde flow to be related to the same mechanism of boundary layer ejection near the critical latitude as the prograde jets described in \cite{noir2009}. To quantify the distance at which the real flow deviates from the analytical solution, we identify the value $r_{min}$ where the velocity has a minimum. The thickness of the layer where the prograde flow develops is plotted in Fig.\ref{fig:graphe_omega}.b and is found to scales as $1/\sqrt{\omega}$, which is representative of a skin effect. So if $\omega$ becomes too small at a fixed $E$, the layer where the effects of external walls are important is visible in the bulk and perturbs the zonal flow.

\section{Conclusion}

In this paper, combining numerical and experimental studies,  we report the first quantitative measurements of the steady flow driven by longitudinal librations in a rotating sphere. This approach confirms the main features of the weakly nonlinear theory of \cite{busse2010}: a retrograde differential rotation induced by the libration of the sphere takes place, which may be assimilated to a solid body rotation for $r < 0.6$. It is also shown that the amplitude of this steady zonal flow is independent of $\omega$ and $E$ and scales as $\epsilon^2$. Note that the same features have been observed experimentally in a librating cylinder \cite[][]{noir2010} and numerically in a librating spherical shell \cite[][]{calkins2010} and thus appear to be generic of librating flows.

The main differences between our results and the theoretical profile of \cite{busse2010} arise close to the outer boundary at the equator. There, we observe a prograde flow in a layer of thickness proportional to $1/\sqrt{\omega}$. The analytical resolution of this peculiar feature would necessitate a special treatment since it appears at the critical latitude of the outer boundary layer \cite[][]{busse2010}. This, as well as the experimental study of a librating spherical shell, will be the subject of forthcoming studies.

\bibliographystyle{jfm}
\bibliography{biblio}

\end{document}